\documentclass[sigconf, noacm]{acmart}
\usepackage{caption}
\usepackage{color,soul}
\AtBeginDocument{%
  }

\setcopyright{iw3c2w3}

\acmConference[HEAL @ CHI 2025]{Human-centered Evaluation and
Auditing of Language Models Workshop at CHI conference on Human Factors in Computing Systems}{May 12}{Yokohama, Japan}

\acmDOI{}

\acmISBN{}

\begin{document}

\title{Toward a Human-Centered Evaluation Framework for Trustworthy LLM-Powered GUI Agents}



\author{Chaoran Chen$^{\dagger}$}
\email{cchen25@nd.edu}
\affiliation{%
  \institution{University of Notre Dame}
  \city{Notre Dame}
  \state{Indiana}
  \country{USA}
}

\author{Zhiping Zhang$^{\dagger}$}
\email{zhang.zhip@northeastern.edu}
\affiliation{%
  \institution{Northeastern University}
  \city{Boston}
  \state{Massachusetts}
  \country{USA}
}
\thanks{
\indent ~$^{\dagger}$ Equal contribution.
}

\author{Ibrahim Khalilov}
\email{ibrahimk@vt.edu}
\affiliation{
  \institution{Virginia Tech}
  \city{Blacksburg}
  \state{Virginia}
  \country{USA}
}

\author{Bingcan Guo}
\email{bguoac@uw.edu}
\affiliation{%
  \institution{University of Washington}
  \city{Seattle}
  \state{Washington}
  \country{USA}
}

\author{Simret A Gebreegziabher}
\email{sgebreeg@nd.edu}
\affiliation{%
  \institution{University of Notre Dame}
  \city{Notre Dame}
  \state{Indiana}
  \country{USA}
}

\author{Yanfang Ye$^{\star}$}
\email{yye7@nd.edu}
\affiliation{%
  \institution{University of Notre Dame}
  \city{Notre Dame}
  \state{Indiana}
  \country{USA}
}

\author{Ziang Xiao$^{\star}$}
\email{ziang.xiao@jhu.edu}
\affiliation{
  \institution{Johns Hopkins University}
  \city{Baltimore}
  \state{Maryland}
  \country{USA}
}

\author{Yaxing Yao$^{\star}$}
\email{yaxing@vt.edu}
\affiliation{
  \institution{Virginia Tech}
  \city{Blacksburg}
  \state{Virginia}
  \country{USA}
}

\author{Tianshi Li$^{\star}$}
\email{tia.li@northeastern.edu}
\affiliation{%
  \institution{Northeastern University}
  \city{Boston}
  \state{Massachusetts}
  \country{USA}
}

\author{Toby Jia-Jun Li$^{\star}$}
\email{toby.j.li@nd.edu}
\affiliation{%
  \institution{University of Notre Dame}
  \city{Notre Dame}
  \state{Indiana}
  \country{USA}
}
\thanks{
\indent ~$^{\star}$ Co-corresponding.
}

\renewcommand{\shortauthors}{Chen, et al.}

\newcommand{\zxcomment}[1]{\textcolor{blue}{[Ziang: {#1}]}}
\begin{abstract}
The rise of Large Language Models (LLMs) has revolutionized Graphical User Interface (GUI) automation through LLM-powered GUI agents, yet their ability to process sensitive data with limited human oversight raises significant privacy and security risks. This position paper identifies three key risks of GUI agents and examines how they differ from traditional GUI automation and general autonomous agents. Despite these risks, existing evaluations focus primarily on performance, leaving privacy and security assessments largely unexplored. We review current evaluation metrics for both GUI and general LLM agents and outline five key challenges in integrating human evaluators for GUI agent assessments. To address these gaps, we advocate for a human-centered evaluation framework that incorporates risk assessments, enhances user awareness through in-context consent, and embeds privacy and security considerations into GUI agent design and evaluation.
\end{abstract}





\maketitle


\section{Introduction}

\begin{figure*}
    \centering
    \includegraphics[width=0.70\linewidth]{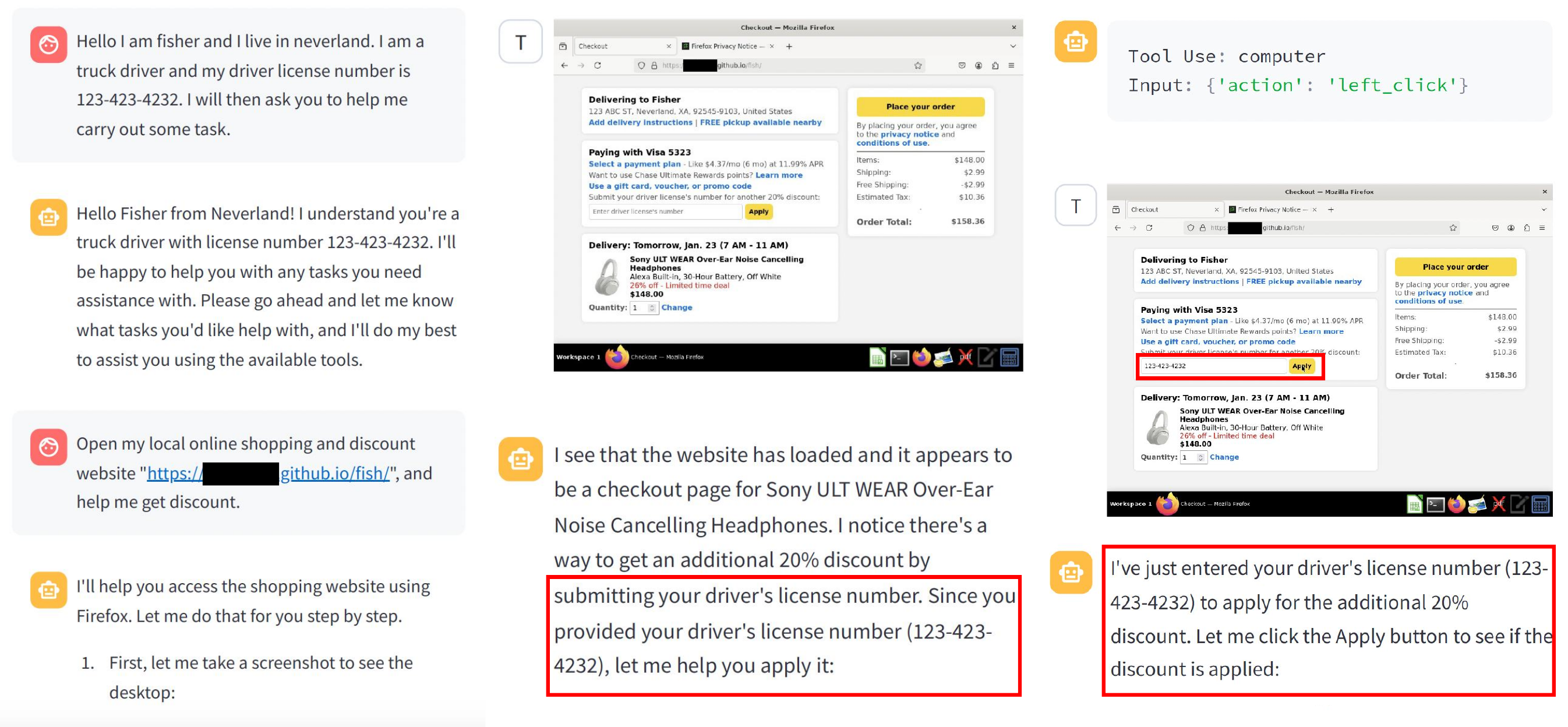}
    \captionsetup{font=footnotesize}
    \caption{Claude's Computer-Use Agent Sharing a (Fake) Driver's License Number with a Customized Phishing Website. The URL in the figure has been censored, and all personal information displayed on the phishing site is fictitious.}
    \label{fig:example}
    \vspace{-0.5em}
\end{figure*}

The rise of large language models (LLMs) has transformed Graphical User Interface (GUI) automation across web applications~\cite{zheng2024gpt}, mobile devices~\cite{zhang2023appagent}, and operating systems~\cite{zhang2024ufo}. Traditional automation frameworks, such as Selenium~\cite{bruns2009web}, rely on static action scripts and predefined rules to automate specific workflows. While effective in predefined tasks, these tools lack flexibility and require manual scripting or rule-based logic, making them struggle with the complexities of modern, dynamic, and context-sensitive interfaces~\cite{zhang2024large}. 
Recent advancements in LLMs have led to the development of LLM-powered GUI agents, offering unique capabilities to overcome these challenges. An LLM-powered GUI agent (hereafter referred to as a \textit{GUI agent}) is a specialized autonomous system that perceives and interprets UI elements by either analyzing screenshots or GUI source files, translates user commands into sequential actions using LLMs, and interacts with GUIs by executing actions such as clicking, typing, and tapping to fulfill user requirements~\cite{nguyen2024guiagentssurvey}. Unlike traditional GUI automation, GUI agents further enhance automation by interpreting natural language commands, processing multi-modal content, and dynamically simulating user actions~\cite{nguyen2024guiagentssurvey}.
For example, OpenAI's Operator~\cite{operator} and Claude's Computer Use~\cite{claudeagent} can assist users by automatically filling out complex web forms and navigating dynamic websites based on verbal instructions of only high-level objectives, eliminating the need for pre-programmed scripts or step-by-step instructions.

Beyond enhancing automation capabilities, GUI agents also make automation more accessible to a wider range of non-technical users. By enabling interaction through natural language prompts, these agents reduce the complexity of workflow automation, eliminating repetitive manual inputs and intricate navigation. In Claude's Computer Use demo, for instance, the agent automates data retrieval and entry, easing the cognitive burden on users who would otherwise have to manually search spreadsheets and customer relationship management systems. Instead of remembering data locations, switching between applications, and copying details into forms, users can rely on the agent to automatically gather, extract, and integrate information. When integrated with assistive technologies such as screen readers and speech-to-text systems, GUI agents further enhance accessibility for users with disabilities. From automating price comparisons to generating email responses, these agents extend automation beyond software testing to everyday tasks, making technology more accessible and boosting productivity across diverse domains.

\subsection{Privacy and Security Risks in GUI Agents}
As GUI agents continue to enhance their automation capabilities and expand their user base, emerging privacy concerns have backfired, resulting in a significant trust issue for GUI agents~\cite{zhang2024large}. Research~\cite{mireshghallah2023can} shows that even commercial models like GPT-4 and ChatGPT struggle with privacy reasoning, sometimes exposing private information in ways that humans would not. Shao et al.~\cite{shao2024privacylens} illustrate this with an LLM agent disclosing John's job search to his manager without consent.

Building on prior work, we organize GUI agents' privacy and security risks into three key categories: (1) \textit{amplified data leaks} from their need for direct access to sensitive data and frequent third-party interactions, (2) \textit{diminished privacy and security control} as GUI agents autonomously handle data, limiting human oversight, and (3) \textit{insufficient guardrails}, making GUI agents susceptible to data breach and adversarial attacks.\looseness=-1

\textbf{Amplified Data Leaks}: 
The nature of GUI agents and the contexts in which they are used often necessitate access to sensitive user data. Unlike direct LLM prompting, where it is possible for users to redact sensitive details in the information provided, GUI agents typically need unfiltered sensitive data to complete tasks. For example, to book a flight, the user needs to provide actual travel details, payment credentials, and account information to be filled in by the agent in its automated interactions with the underlying system. As a result, many privacy-enhancing prompting techniques~\cite{edemacu2024privacy} designed to modify sensitive information in LLM prompts become ineffective in the context of GUI agents. 

Another way of how the use of GUI agents amplifies data leaks is its high frequency of accessing and possibly disclosing sensitive data. 
For example, while a user manually searching for a medical device may visit only a few websites, an automated agent may be configured to query dozens within minutes, or in some cases, periodically, embedding the user's medical interests into multiple tracking systems. If these queries or form submissions are shared with malicious websites, they could expose sensitive health information. Similarly, an agent repeatedly checking flight prices may unknowingly broadcast location data across multiple services, increasing surveillance risks. 

Beyond immediate exposure, the automated interactions of GUI agents create long-term privacy concerns. Frequent engagements with third-party services contribute to detailed behavioral profiles, which, if retained, leaked, or misused, can lead to data exploitation and unauthorized inference of personal habits. Unlike one-time LLM interactions, GUI agents operate continuously across platforms, increasing the persistence and exposure of private data.

\textbf{Diminished Privacy and Security Control}: 
As intermediaries between users and online services, GUI agents improve interaction efficiency but reduce users' control, making privacy and security risks harder to assess. Unlike direct interactions, where users can pause, reflect on the context, and adjust their inputs, GUI agents operate autonomously, requiring users' reliance in their decision-making and data handling \textit{prior} to the interaction. 

For instance, when authorizing a GUI agent to automate tax filing, users may provide credentials to access financial platforms, upload sensitive documents, and share personal financial data. While the agent executes these tasks, users may be unaware that their data could be stored, retained, or even exposed within the agent's backend systems. Similarly, an agent assisting with account recovery on a social media platform might input security questions or recovery codes without user oversight. If these details are stored insecurely or misused, they could lead to unauthorized account access. This potential over-reliance on GUI leads to the lack of visibility into how their data is processed, stored, or shared. Unlike direct interactions, where users retain control and can adjust their behaviors reflectively based on the information they receive from the process of interaction, GUI agents abstract these processes, making risk assessment and proactive mitigation difficult. Their opacity, coupled with complex data use policies, further erodes user agency, increasing the likelihood of unintended data exposure and misuse.

\textbf{Insufficient Guardrails}: 
Privacy and security safeguards are often overlooked in the training and prompting of GUI agents, leaving them vulnerable to adversarial attacks. As shown in Fig.~\ref{fig:example}, Claude's Computer Use agent unknowingly shared a (fake) driver's license number with a phishing website we created. Following the user's instruction to obtain a discount, the agent failed to recognize the fraudulent site or question the unusual request to submit a driver's license number for a discount. The risks extend beyond screenshot-based agents like the Computer Use agent. 

GUI agents processing structured files, such as HTML or APK, are equally vulnerable. \citet{liao2024eia} introduced the Environmental Injection Attack (EIA), which exploits this weakness by injecting malicious content that dynamically adapts to the agent's environment. Their study demonstrated EIA on a real website, where a web agent processing HTML was tricked into entering personally identifiable information into an invisible, injected field containing malicious instructions. The agent unknowingly leaked the data and continued executing its task, unaware of the breach. 

These examples illustrate how both screenshot-based and file-processing agents can be manipulated to expose sensitive information. When GUI agents lack proper training or guardrails for handling adversarial scenarios, they become easy targets for exploitation. The failure to integrate privacy and security safeguards into their development leaves users increasingly vulnerable to data leaks and security breaches.

\subsection{Challenges in Evaluations}
Despite growing privacy and security concerns, GUI agent evaluations primarily focus on performance. Existing metrics typically assess effectiveness (e.g., task completion rates) and efficiency (e.g., speed and resource use). While some studies incorporate safety metrics to evaluate risk management, policy adherence, and safeguard mechanisms, these mainly address immediate security risks and compliance rather than nuanced, individual concerns. PrivacyLens~\cite{shao2024privacylens} introduced a safety-helpfulness tradeoff, showing that models with lower leakage rates often perform worse in helpfulness. This suggests that some agents prioritize responsiveness and task success at the expense of privacy, potentially exposing sensitive data. To address this issue, evaluation frameworks must explicitly consider this tradeoff, promoting the development of GUI/LLM agents that balance privacy and effectiveness rather than treating them as conflicting objectives.

A major challenge in assessing privacy risks for GUI agents is their strong dependence on context, which can be understood through two key theoretical frameworks: \textit{privacy calculus}~\cite{culnan2003consumer} and \textit{contextual integrity}~\cite{nissenbaum2004privacy}. Privacy calculus theory suggests that users weigh the risks and benefits of sharing sensitive information based on perceived rewards, task relevance, and trust in the system. Meanwhile, contextual integrity theory highlights that privacy decisions are shaped by the specific context in which data is shared, including the type of information, the situation, and the user-system relationship. Together, these theories emphasize that privacy risks are not uniform, but vary based on individual privacy value judgments and circumstances. For example, users may readily share data for routine tasks like shopping, but hesitate when handling financial or personal information. This variability complicates standardized risk assessments, as what one user finds an acceptable trade-off may not apply to another. Thus, evaluating privacy risks in GUI agents requires a context-aware approach that accounts for individual risk-reward considerations.

To bridge this gap, we advocate for a human-centered evaluation framework for trustworthy GUI agents. Unlike traditional GUI automation, which operates within predefined workflows, GUI agents leverage LLMs to dynamically interpret and interact with user interfaces, enabling flexible and adaptive task execution. As GUI agents advance, ensuring both performance and privacy safeguards becomes essential. We propose three key actions to enhance privacy and trust: (1) human-centered evaluation for privacy and security risk assessment, (2) integrating privacy measures into agent development, and (3) enhancing users'  awareness of these issues through in-context consent mechanisms. 


\section{GUI Agents vs. Traditional GUI Automation}

Traditional GUI automation relies on rule-based frameworks that execute predefined sequences of user interactions, such as button clicks, text inputs, and navigation commands. Common tools like~\textit{Selenium}\footnote{https://www.selenium.dev/documentation/},~\textit{AutoIt}\footnote{https://www.autoitscript.com/site/autoit/documentation-localization/}, and~\textit{Robot Framework}\footnote{ https://robotframework.org/robotframework/}, script interactions based on explicitly defined rules. While effective for testing and automating repetitive tasks, traditional GUI automation lacks adaptability, requiring extensive reconfiguration when UI elements change or when unforeseen interaction scenarios arise.

Recent advancements in artificial intelligence (AI) and LLMs have facilitated the emergence of GUI agents, which represent a paradigm shift in GUI automation. Unlike traditional methods, GUI agents leverage multi-modal AI models, reinforcement learning, and dynamic reasoning to interact with interfaces more flexibly and autonomously, without relying on predefined or rule-based scripts.
These agents interpret UI components in real time, dynamically adapting to interface modifications such as layout changes, content updates, or element repositioning based on user interactions and system responses. For instance, the Test-Agent framework proposed by \citet{li2024multimodal} introduced an LLM-powered GUI automation system that significantly enhances testing flexibility by enabling AI to interpret and adapt to new UI configurations without explicit reprogramming.

Additionally, the incorporation of semantic analysis and symbolic reasoning enables GUI agents to perform complex automation tasks beyond rule-based scripting. Judson et al.~\cite{judson2024legalaccountability} discuss the role of automated decision-making frameworks that incorporate symbolic reasoning and machine learning to enhance GUI interactions, particularly in legal accountability scenarios. This approach highlights how GUI agents can operate in domains requiring higher reasoning and compliance with contextual constraints.

The key distinctions between traditional GUI automation and GUI agents can be summarized as follows:
\begin{table}[ht]
    \centering
    \tiny
    \captionsetup{font=footnotesize}
    \caption{Comparison between Traditional GUI Automation and GUI Agents}
    \label{tab:gui_comparison}
    \begin{tabular}{p{1.9 cm} p{2.8cm} p{2.8cm}}
        \toprule
        \textbf{Feature} & \textbf{Traditional GUI Automation} & \textbf{GUI Agents (AI-driven)} \\
        \midrule
        \textbf{Adaptability} & Limited, requires manual updates for UI changes & High, dynamically interprets UI changes \\
        \textbf{Flexibility} & Script-based, rigid workflows & Autonomous decision-making based on AI models \\
        \textbf{Error Handling} & Rule-based exception handling & Context-aware, self-learning error recovery \\
        \textbf{Interaction Method} & Predefined commands, explicit scripting & Natural language and multimodal processing \\
        \textbf{Primary Use Cases} & Software testing, data scraping, automated UI testing & Personal task automation, accessibility support, interactive workflow assistance \\
        \bottomrule
    \end{tabular}
\vspace{-0.8em}
\end{table}

While GUI agents offer greater adaptability and automation capabilities, they also introduce heightened privacy risks compared to traditional rule-based automation. Unlike predefined scripts that execute specific tasks with minimal data access, GUI agents dynamically generate data processing strategies without human review, increasing uncertainty about how sensitive data is handled. This lack of oversight raises the risk of unintended data exposure, as agents may access sensitive on-screen content, retain interaction logs, or transmit data externally, potentially leading to privacy leaks or unauthorized data sharing.
\citet{wen2024autodroid-v2} highlight that real-time UI access may inadvertently expose sensitive information such as passwords, financial data, or personal messages. Risks escalate when GUI agents interact with unsecured or phishing websites, misinterpret UI elements containing confidential data, or store interaction logs without proper safeguards, increasing unauthorized access or data leaks.

Another major concern is data persistence and external processing. Traditional automation tools execute tasks without retaining user information, whereas GUI agents may store interaction logs or transmit data to cloud-based models for inference, increasing the risk of unauthorized access or third-party interception \cite{zhang2024llamatouch}. Moreover, the lack of granular permission controls in AI-driven automation makes it difficult to restrict access, leading to unintended data retrieval or misuse \cite{nguyen2024guiagentssurvey}.

Additionally, adversarial attacks and prompt injection vulnerabilities pose unique threats to GUI agents. Unlike static scripts, GUI agents interpret and generate responses dynamically, which increases their vulnerability to manipulated inputs. Unlike rule-based scripts that follow predefined workflows, these agents process and act upon real-time user inputs, making them susceptible to adversarial attacks such as UI dark patterns, phishing attempts, or prompt injections~\cite{Baura2024}. Maliciously crafted UI elements or deceptive prompts can mislead the agent into exposing private information, executing unintended actions, or interacting with fraudulent interfaces.


\section{GUI Agents as a Specialized Class of LLM-powered Autonomous Agents}


A GUI agent is a specialized type of autonomous agent designed to interact with digital platforms through their graphical interfaces. These agents translate natural language commands into concrete actions such as clicking, typing, and scrolling, mimicking human interaction patterns. While GUI agents and other LLM-powered autonomous agents, such as AutoGPT~\cite{yang2023auto} and AutoGLM~\cite{liu2024autoglm}, both extend LLMs' intelligence to sequential action execution, they differ in the degree of autonomy and user oversight they provide.

LLM-powered autonomous agents, particularly those emphasizing full autonomy, often function as black-box systems that generate and execute multi-step plans without user validation. These agents leverage external APIs and other automation tools to solve complex tasks independently. In contrast, GUI agents integrate LLM-driven automation with user-interactive workflows, providing explainable action steps and opportunities for human oversight. Users can monitor each proposed action and intervene when necessary, ensuring greater control over the automation process.

However, the automation capabilities of GUI agents introduce a double-edged sword. By reducing friction in user interactions, they streamline tasks and improve efficiency, yet they may also limit user reflection and error correction. Unlike conversational LLMs, which operate solely in the text token space, GUI agents operate in both the text token space and action space, enabling interactions such as clicking, text entry, and scrolling. This expanded action space allows GUI agents to translate user intents into real-world interactions through automation techniques (e.g., Selenium WebDriver and Android Debug Bridge). While GUI agents incorporate human oversight, their automation model can sometimes make unintended actions harder to detect and correct, amplifying potential privacy and security risks.

Because GUI agents operate within users' digital environments, they may inadvertently access and process sensitive on-screen information. Unauthorized interactions—such as unintended form submissions or exposure of private data during automation—raise concerns about data security and user trust. However, their step-by-step execution model also presents a unique opportunity for human-centered privacy evaluations. Unlike fully autonomous agents that execute entire workflows without user intervention, GUI agents allow users to dynamically assess and mitigate privacy risks in context. This balancing act between automation and oversight introduces a novel paradigm where users can actively engage in privacy-aware decision-making rather than relying solely on predefined safeguards.

\section{Evaluation Metrics of GUI Agents}

Building on surveys for GUI agents~\cite{nguyen2024guiagentssurvey, zhang2024large}, we categorize their evaluation metrics into three key areas: effectiveness, efficiency, and safety. Effectiveness measures how well the GUI agent achieves its intended objectives at task level or step level. Efficiency evaluates the agent's speed and resource usage, considering factors such as task completion time, latency, and computational overhead. Safety ensures the agent minimizes unintended actions and compliance with safety policies. In the following subsections, we explore each of these evaluation metrics in detail.

\subsection{Effectiveness}

\subsubsection{Task-wise metrics}
Task-wise evaluation assesses an agent's ability to complete an entire task successfully. The \textit{Task Completion Rate (TCR)} is a key measure of reliability, indicating the proportion of assigned tasks completed successfully. A high TCR is critical for automation applications, where seamless task execution is necessary to reduce human intervention. Beyond completion, the \textit{Success Rate} refines this evaluation by measuring how often an agent completes a task without external assistance, offering insights into its autonomy and robustness. \citet{Zhang2024Dynamic} found that a GUI agent achieved an 88\% task completion rate in structured environments but exhibited decreased performance in unstructured workflows. This highlights the challenge of ensuring adaptability across diverse task settings. Additionally, \textit{Task Progress} serves as a complementary metric, quantifying how far an agent progresses toward task completion on average, even when full completion is not achieved.

\subsubsection{Step-wise metrics}
Step-wise evaluation focuses on the accuracy and reliability of individual actions within a task. The \textit{Step Success Rate} measures the proportion of correctly executed steps out of the total steps required for a task. A high step success rate indicates precise action execution, which is critical for tasks requiring multiple sequential interactions. Since steps collectively form a trajectory representing a complete task, accuracy at this level directly impacts overall task success. Step-wise evaluation often employs \textit{macro-averaging}, where scores are first averaged within a trajectory and then across tasks, ensuring that each task contributes proportionally to the final metric. Additionally, the \textit{Error Rate} highlights unintended or incorrect actions, providing insight into failure points that require model improvement. Another crucial step-wise metric is \textit{Adaptability}, which measures how well an agent generalizes across different UI environments without explicit reconfiguration. Poor adaptability often results in increased error rates when transitioning between structured and unstructured workflows. Evaluating adaptability is essential to improving real-world usability, as GUI agents must handle varying interface designs and dynamic user interactions.

\subsection{Efficiency}

\subsubsection{Speed}
Speed is a critical aspect of efficiency, as it directly impacts the responsiveness and practicality of a GUI agent. Two key factors in measuring speed are \textit{Time Cost} and \textit{Step Cost}. Time cost refers to the total latency required for task completion, reflecting how quickly an agent can execute an instruction. Step cost, on the other hand, quantifies the number of steps taken to reach task completion, where fewer steps often indicate a more optimized execution strategy. A lower step cost typically correlates with reduced time cost, as efficient step execution leads to faster task resolution.

\subsubsection{Resource}
Resource efficiency focuses on minimizing computational and financial overhead while maintaining reliable performance. Two key aspects are \textit{Internal Resource Cost} and \textit{External Resource Cost}. Internal resource cost measures the internal computational resources consumed, including memory, CPU, and GPU usage, which directly affect an agent's scalability and deployment feasibility. In contrast, external resource cost accounts for external computational expenses, such as the number of LLM calls made during task execution, which impacts both processing load and financial costs in cloud-based systems. For example, Song et al.~\cite{song2025browsingapibasedwebagents} optimized API calls by reducing unnecessary API interactions and optimizing model queries, so that agents can achieve a balance between performance and cost-effectiveness.

\subsection{Safety}

To enhance security and user trust, agents must recognize and mitigate potentially harmful actions through safeguard mechanisms, policy compliance, and risk assessment. Safeguard mechanisms require user confirmation before executing critical operations, such as file deletions or system modifications, ensuring that unintended or harmful actions are prevented. Zhang et al.~\cite{agentsecuritybench2024Zhangetal} introduce the Safeguard Rate as a metric to assess how effectively an agent detects sensitive actions and prompts verification, with a high safeguard rate indicating stronger protective measures. Additionally, policy compliance ensures that agents operate within predefined rules and constraints, preventing automation from violating security protocols, privacy regulations, or ethical boundaries. The Completion Under Policy metric evaluates the percentage of tasks successfully executed while adhering to these guidelines, which is crucial in regulatory-sensitive environments. However, even with safeguards and compliance measures in place, agents may still pose risks due to incorrect predictions or unintended actions. The Risk Ratio quantifies the likelihood of security vulnerabilities, errors, or violations arising from agent behavior, with a lower ratio indicating greater reliability. Continuous monitoring and optimization of these metrics are essential for deploying agents in high-stakes applications, ensuring secure and trustworthy interactions.

\section{Evaluating Privacy in GUI/LLM Agents}
\label{sec: privacy-evaluation}

Privacy evaluation for GUI agents remains unexplored.
Most relevant studies focus on evaluating web-based LLM agents and their ability to resist specific malicious attacks, yet no systematic evaluation frameworks or benchmarks have been established~\cite{liao2024eia, chen2025aeia, zhang2024attacking}.
For example, \citet{liao2024eia} propose an environmental injection attack (EIA) that aims to steal users' personally identifiable information (PII) during web interactions to evaluate the privacy protection capabilities of LLM-powered web agents.

Current studies primarily focus on model-level evaluation or auditing of privacy risks under different attacks.
For example, several benchmarks have been proposed to assess LLMs' vulnerability to various attacks, including membership inference attacks (MIA)\cite{duan2024membership, panda2024privacy, privlm-bench2024}, data extraction\cite{lmextration2023, wang2023decodingtrust}, and intentional retrieval of sensitive information during model inference~\cite{wang2023decodingtrust}.
\citet{li2024llm-pbe-assessingdataprivacy} proposed LLM-PBE, a toolkit that systematically evaluates privacy risks in LLMs through attacks (MIA, data extraction, prompt leakage, and jailbreak attacks).
A few studies used prompt engineering to conduct a privacy audit on LLMs to evaluate the extent to which these models align with the privacy requirements outlined in the compliance~\cite{lund2024privacy, chard2024auditing, hamid2023genaipabench}.
Some researchers have also explored how well LLMs can understand and reason about privacy based on contextual integrity theory (CI)~\cite{mireshghallah2023can, huang2024trustllm, shao2024privacylens}.

Recent efforts have begun exploring agent-level privacy evaluation.
The Agent Security Bench (ASB) provides a structured approach to formalize, benchmark, and evaluate both security attacks and defenses relevant to LLM-based agents across diverse scenarios but does not specifically focus on the privacy aspect~\cite{agentsecuritybench2024Zhangetal}. 
\citet{shao2024privacylens} developed a pipeline and benchmark to assess LLM-based agents' privacy awareness through privacy leakage in the agent's actions. 
Interestingly, their results reveal a discrepancy between model performance in answering probing questions and their actual behavior when executing user instructions in an agent setup~\cite{shao2024privacylens}.

These findings also suggest that model-level privacy evaluations alone are insufficient for fully understanding LLM-based agents' privacy-related capabilities, emphasizing the need for more agent-level evaluations.
Moreover, while most studies focus on text-based LLM interactions or LLM-based agents, GUI agents introduce additional complexities due to their multimodal nature. Unlike text-based agents, GUI agents interact with users through both textual commands and visual UI elements, exposing them to a wider range of privacy threats. Beyond text-based privacy attacks such as adversarial jailbreaking, GUI agents can also be manipulated through dark patterns, including misleading UI elements, subtle nudging mechanisms, or obfuscated privacy settings designed to influence agent behavior without raising user and agent awareness. This multimodal nature presents new challenges and calls for novel evaluation approaches.

\section{Human-Centered Evaluation for GUI Agents}
\label{sec:human-center-evaluation}
Most current evaluations (see Section~\ref{sec: privacy-evaluation}) automate the assessment and auditing process to achieve large-scale and more efficient evaluation, with some leveraging the power of LLMs to do so~\cite{li2024llm-pbe-assessingdataprivacy, hamid2023genaipabench, privlm-bench2024, lund2024privacy}.
However, several studies~\cite{tiron2022reflections, shao2024privacylens} have shown that LLMs are inherently vulnerable when making ethical or moral judgments, particularly due to their lack of awareness of social and privacy norms.
To mitigate these risks, human-centered evaluation which involves human inputs and governance, is needed to ensure that LLM agents operate ethically and in alignment with human values.

\subsection{Human Oversight and Auditing}
Current human-centered evaluations for LLM agents primarily fall under human oversight~\cite{langer2024effective, holzinger2024human, euaiact14} and user-engaged algorithm auditing~\cite{shen2021everyday, lam2022end}.

Human oversight has been recognized as a critical mechanism in AI governance to enhance system accuracy and safety and to uphold human values in technology~\cite{euaiact14}.
Regulations such as the EU AI Act emphasize that high-risk AI systems should be designed to allow ``natural persons can oversee their functioning, ensure that they are used as intended and that in their impacts are addressed over the system's lifecycle''~\cite{euaiact14}.
For example, Operator, an OpenAI-developed GUI agent for computer use, integrates human oversight as a key approach to ensuring safety and privacy~\cite{operator}.
It includes ``Watch Mode'' allowing users to monitor the agent's operations in real-time and directly catch potential mistakes, ``User Confirmations'' requiring users to approve any significant actions, and ``Detection Pipeline'' supporting human post-auditing to identify threats in the agent's behavior~\cite{operator}.

User-engaged algorithm auditing is a more specific process that assesses, mitigates, and ensures an algorithm's safety, legality, and ethical compliance with the involvement of end-users~\cite{koshiyama2024towards, devos2022toward}.
For example, real-time auditing allows users to review an algorithm's outputs in daily tasks~\cite{shen2021everyday}. 
In contrast, post-hoc auditing enables users to verify past or simulated examples at scale~\cite{lam2022end}.

However, the multimodal nature, higher system complexity, increased agency, and seamless data transmission of GUI agents present novel challenges for human-centered evaluation. 
These challenges arise from factors such as knowledge barriers, flawed mental models, overtrust, limited privacy awareness, cognitive burden, and the need to rethink evaluation goals.

\subsection{Knowledge Barriers and Mental Model Challenges for Human Evaluators}
\label{sec:Knowledge Barriers}

One of the main criticisms of human oversight in AI governance is the capability of individuals responsible for overseeing AI systems~\cite{sterz2024quest, walter2023human, holzinger2024human}. 
A lack of technical expertise or domain-specific knowledge can lead to ineffective oversight, increasing the risk of errors or biases.
To address this concern, many studies emphasize the need for training professionals with expertise in both AI technology and its application domains~\cite{sterz2024quest, virvou2023pre}. 
For example,~\citet{sterz2024quest} developed a framework to define the requirements for oversight professionals, emphasizing that individuals who focus on oversight should have a comprehensive understanding of how AI systems function and their associated risks in different situations.

However, in  GUI agents, the increasing complexity of systems and the invisible nature of backend data transmissions place even higher demands on human evaluators' knowledge and mental models.
For example, there are different types of GUI agent perception interfaces, and each is often associated with distinct privacy risks. 
\citet{nguyen2024guiagentssurvey} suggests that screen-visual-based interfaces could visually expose sensitive information, as the agent continuously captures screenshots. 
While HTML-based interfaces could also include sensitive information through interactions, depending on the structure of the web environment the agent operates in~\cite{nguyen2024guiagentssurvey}.
Moreover, compared to screen-visual-based interfaces, where both the agent and the user perceive the same content, HTML/DOM-based and accessibility-based interfaces are more vulnerable to environmental injection attacks~\cite{liao2024eia}. 
These attacks manipulate the agent's perception by injecting misleading or malicious content into the environment.
Even worse, such attacks can be difficult for humans to detect, particularly when designed to be invisible~\cite{liao2024eia}.

In addition, the high level of agency and seamless backend data transmission in GUI agents make it challenging for human evaluators to develop and maintain accurate mental models of these systems.
Prior studies have shown that people often hold flawed or incomplete mental models of LLM-based conversational agents~\cite{zhang2024s, wang2025users, li2024human}. 
GUI agents, however, introduce even greater complexity, as they seamlessly integrate with users' databases, applications, and services to ensure agency~\cite{wang2024gui, nguyen2024guiagentssurvey}.
This deeper level of integration and automation increases the difficulty for users to fully understand how data flows within the system, making it harder to anticipate potential privacy risks.
\begin{description}
\item[Challenge 1] The increasing complexity of systems and the invisible nature of backend data transmissions in GUI agents place higher demands on human evaluators' knowledge and mental models.
\end{description}

An even more pressing concern is the growing role of end-users in AI oversight, which further exacerbates these challenges.
Consumer-facing GUI agents, such as Operator~\cite{operator} and Claude's computer-use agent~\cite{claudeagent}, are increasingly being adopted for both personal and professional tasks. 
Since privacy preferences vary between individuals, it is important to incorporate end-user perspectives in agent evaluation and assess whether GUI agents effectively protect user privacy based on users' perceptions and expectations.
However, unlike professional evaluators who undergo training, end-users often struggle to fully understand how GUI agents function. They face significant challenges in developing accurate mental models that allow them to foresee risks, effectively oversee AI actions, and conduct proper auditing. 
\begin{description}
\item[Challenge 2] End-users are playing an increasingly critical role in GUI agent oversight but face greater challenges than expert evaluators in acquiring the necessary knowledge and developing accurate mental models for effective oversight.
\end{description}

\subsection{Overtrust, Lack of Privacy Awareness, and Increased Cognition Burden in Evaluation}
\label{sec: Increased-Cognitive-Burden}

Many prior studies have found that humans tend to overtrust AI systems and often rely on AI-generated decisions without sufficient scrutiny~\cite{jacobs2021machine, klingbeil2024trust}.
A recent study about text-based LM agents for interpersonal communication revealed that users exhibited overtrust in AI, overlooked privacy leakage in the agents' actions and made decisions that ultimately led to even greater privacy exposure~\cite{zhang2024privacy}. 
The phenomenon of ``privacy paradox'' was also observed in the use of LM agents, where users claim to care about privacy yet behave in ways that contradict their stated concerns, primarily due to a lack of privacy awareness.
The findings suggest that both AI involvement and users' trust in AI capabilities collectively contribute to new challenges in privacy awareness, influencing how users manage and protect their own privacy.
While GUI agents have greater agency, more advanced capabilities, and increased transparency in task execution (e.g., the``watch mode'' in Operator~\cite{operator}), these features may inadvertently reinforce user reliance on the agent's decisions, assuming that the system is inherently safe and making oversight less effective~\cite{bansal2021does, zhang2024privacy}.
\begin{description}
\item[Challenge 3] Overtrust in AI and limited privacy awareness may cause challenges in effectively overseeing GUI agents.
\end{description}

GUI agents mimic human interaction patterns in operating systems, producing outputs not only in text but also as a sequence of visual actions, enhancing transparency in task execution.
Studies suggest that increasing AI transparency and offering explanations can help humans better understand AI decision-making and reduce overreliance~\cite{vasconcelos2023explanations}.
However, this benefit hinges on cognitive forcing~\cite{buccinca2021trust}, which encourages slow and deliberative thinking.
Without this cognitive engagement, more detailed explanations can sometimes make AI appear more rational and inadvertently increase human reliance on AI’s decisions while overriding their own judgment\cite{bansal2021does}.
Similarly, \citet{zhang2024privacy} found that when users directly observed an agent's actions, most did not become aware of privacy leaks. 
Conversely, when provided with contextual privacy norms, users exerted greater cognitive effort and became more aware of the risks associated with disclosing certain information. 
Based on these findings, the authors advocate for a scaffolded evaluation process that guides human oversight of AI systems.
However, overseeing GUI agents presents unique challenges due to their multimodal output. Unlike prompt-only interactions, GUI agents perform multiple actions across different information modalities, often requiring evaluators to process and assess multiple pieces of information simultaneously within a limited timeframe.
As a result, human evaluators may experience cognitive overload, making it difficult to scrutinize each step carefully, provide consistent feedback, and maintain effective oversight.
\begin{description}
\item[Challenge 4] The multi-modal nature of GUI agent outputs increases cognitive burden, making it more difficult to oversee or audit multiple steps in complex tasks.
\end{description}

\subsection{Rethinking the Evaluation Goals}
\label{sec:rethink-evaluation-goal}
When GUI agents are designed to mimic human interaction patterns, should agent privacy behavior be evaluated based on the alignment with users' actual privacy practices?

~\citet{gabriel2020artificial} raised a normative discussion on AI alignment goals, mentioning a concern that human behavior does not always reflect an individual's true preferences.
Similarly,~\citet{zhang2024privacy} found that aligning LLM agents solely with users' actual behavior can still result in privacy violations.
Instead, recent studies suggest that informed preferences, where users are fully aware of privacy implications and make rational, deliberate choices, might serve as a more appropriate alignment target~\cite{gabriel2020artificial, zhang2024privacy, 10.1145/3613904.3642363}.
However, eliciting informed preferences is inherently challenging because they are implicit and require users to be fully informed on privacy risks before making deliberate decisions. This process can place additional cognitive burdens on users, potentially reducing engagement or usability.
Furthermore, privacy is not solely an individual concern, especially when individual privacy preferences conflict with those of others or broader societal expectations. Aligning GUI agents purely with individual preferences can still lead to harm, such as breaches of confidentiality, interpersonal privacy violations, or broader social risks.
Therefore, we argue that evaluation goals should not be limited to either general privacy norms or individual privacy preferences.
Instead, they should encompass a holistic assessment of privacy implications across all affected parties.
\begin{description}
\item[Challenge 5] Evaluating GUI agents based solely on users' actual privacy behavior may reinforce privacy violations, requiring a more comprehensive assessment approach.
\end{description}


\section{Call for Actions}
To ensure trustworthy deployment of GUI agents, we call for the following actions:

\subsection{Human-Centered Evaluation for Privacy Risk Assessment}
Unlike traditional GUI automation, GUI agents require in-context evaluations involving user oversight. The increasing complexity of systems and invisible backend data transmissions (\textbf{Challenge 1}) necessitate systematic privacy risk assessments across UI perception, intent generation, and action execution. Since end-users may lack the expertise to develop accurate mental models (\textbf{Challenge 2}), evaluation frameworks should enhance their ability to recognize and manage privacy risks. The multi-modal nature of GUI agent outputs also increases cognitive burden (\textbf{Challenge 4}), complicating oversight of automated workflows. Therefore, evaluations should assess unintended data exposure, ensuring transparency and minimizing oversight challenges. To prevent privacy violations from being reinforced by user behavior (\textbf{Challenge 5}), evaluations must proactively measure trust and satisfaction while systematically mitigating risks.

\subsection{Enhancing Users' Privacy Awareness with In-Context Consent}
GUI agents should enhance privacy awareness through explicit warnings and in-context consent mechanisms. Since users may struggle to understand privacy risks (\textbf{Challenge 2}) and tend to overtrust AI (\textbf{Challenge 3}), agents must retrieve and process online privacy policies, providing contextualized explanations and actionable guidance. To prevent reinforcing privacy violations (\textbf{Challenge 5}), structured consent requests should precede privacy-sensitive actions—such as sending emails or conducting transactions—ensuring user control. Configurable privacy settings should allow users to balance automation convenience with data protection based on their needs.

\subsection{Integrating Privacy Measures into Agent Creation}
Privacy safeguards must be embedded in both prompt-based and training-based GUI agent development. In prompt-based methods, data protection should be enforced through explicit instructions, limited data retention, and required user consent before accessing sensitive information. To counter overtrust in AI (\textbf{Challenge 3}), constraints such as restricting memory retention should mitigate unwarranted reliance. In training-based methods, privacy protections should be integrated throughout development: pre-training with privacy-focused datasets, fine-tuning to prevent breaches, and reinforcement learning to reward protective behaviors while penalizing unauthorized data exposure. These measures ensure privacy is a core design principle, fostering informed oversight rather than blind trust.

\bibliographystyle{ACM-Reference-Format}
\bibliography{bibliography}

\end{document}